\documentclass[11.5pt, twocolumn]{article}
\usepackage[a4paper, total={6.5in, 9.5in}]{geometry}
\pdfoutput=1

\usepackage{graphicx}
\usepackage{dcolumn}   
\usepackage{bm}        
\usepackage{amssymb}   
\usepackage{float}
\usepackage{xcolor}
\usepackage{soul}
\usepackage[normalem]{ulem}
\usepackage{authblk}

\newcommand{\strikeout}[1]{}

\begin{document}

\title{Non-linear extended MHD simulations of type-I edge localised mode cycles in ASDEX Upgrade and their underlying triggering mechanism}

\author[1]{\large A.~Cathey\thanks{andres.cathey@ipp.mpg.de}} 
\author[1]{\large M.~Hoelzl}
\author[1]{\large K.~Lackner}
\author[2,3]{\large G.T.A.~Huijsmans} 
\author[1]{\large M.G.~Dunne}
\author[1]{\large E.~Wolfrum}
\author[4]{\large S.J.P.~Pamela} 
\author[5]{\large F.~Orain} 
\author[1]{\large S.~G\"unter}
\author[6]{\large the JOREK team} 
\author[7]{\large the ASDEX Upgrade Team} 
\author[8]{\large the EUROfusion MST1 Team} 

\affil[1]{\small Max Planck Institute for Plasma Physics, Boltzmannstr.2, 85748 Garching, Germany}
\affil[2]{\small CEA, IRFM, 13108 Saint-Paul-Lez-Durance, France}
\affil[3]{\small Eindhoven University of Technology, P.O. Box 513, 5600 MB Eindhoven, The Netherlands}
\affil[4]{\small CCFE, Culham Science Centre, Abingdon, Oxon, OX14 3DB, United Kingdom}
\affil[5]{\small Centre de Physique Théorique, Ecole Polytechnique, CNRS, France}
\affil[6]{\small see https//www.jorek.eu for a list of current team members}
\affil[7]{\small see the author list of H. Meyer et al. 2019 Nucl. Fusion 59 112014}
\affil[8]{\small see the author list of B. Labit et al. 2019 Nucl. Fusion 59  0860020}

\date{}
\maketitle

\begin{abstract}
A triggering mechanism responsible for the explosive onset of edge localised modes (ELMs) in fusion plasmas is identified by performing, for the first time, non-linear magnetohydrodynamic simulations of repetitive type-I ELMs. Briefly prior to the ELM crash, destabilising and stabilising terms are affected at different timescales by an increasingly ergodic magnetic field caused by non-linear interactions between the axisymmetric background plasma and growing non-axisymmetric perturbations. The separation of timescales prompts the explosive, i.e. faster than exponential, growth of an ELM crash which lasts ${\sim 500~\mathrm{\mu s}}$. The duration and size of the simulated ELM crashes compare qualitatively well with type-I ELMs in ASDEX Upgrade. As expected for type-I ELMs, a direct proportionality between the heating power in the simulations and the ELM repetition frequency is obtained. The simulations presented here are a major step forward towards predictive modelling of ELMs and of the assessment of mitigation techniques in ITER and other future tokamaks.
\end{abstract}


\section{Introduction} 
High-confinement mode (H-mode)~\cite{wagner1982regime} defines the standard operational scenario to achieve power amplification in ITER~\cite{iter1999mhd}. This operational regime hosts a steep pressure profile in the edge of the confined region which, in turn, drives a large toroidal current. Under such conditions, magnetohydrodynamic (MHD) instabilities called \textit{edge localised modes} (ELM) can become excited and rapidly (within $\sim0.1$ to $1~\mathrm{ms}$) eject hot plasma towards the plasma facing components~\cite{zohm1996edge,doyle1991modifications,huysmans2005elms,kirk2006evolution}. The steep edge pressure profile together with the large toroidal current crash as a result, but begin to gradually recover until the process repeats itself, thus defining an ELM cycle. Type-I~ELMs, the most pernicious type of such instabilities, repetitively expel between $5\%$ and $15\%$ of the plasma stored energy to the material surfaces. The associated heat fluxes pose significant concerns for next-step devices like ITER and must be completely avoided in a future reactor~\cite{eich2017elm}.

Resulting from the destructive potential inherent to type-I~ELMs, and in order to produce physics-based predictions for future machines, substantial effort has been dedicated from experiment and theory to understand the underlying mechanisms that drive and trigger these instabilities~\cite{zohm1996edge,doyle1991modifications,huysmans2005elms,leonard2014edge,wilson2004theory,huijsmans2015modelling}. Non-linear MHD simulations of ELMs in realistic tokamak geometry with various codes have played an increasingly important role in this regard~\cite{sugiyama2010magnetic,sugiyama2012intrinsic,ferraro2010ideal,snyder2005progress,xu2011nonlinear,xi2014phase,xu2019promising,king2016impact,huysmans2009non,huijsmans2013non,orain2013non,hoelzl2018insights,futatani2014non,becoulet2014mechanism,becoulet2017non,pamela2015non}. However, the simulations performed so far have had the shortcoming of modelling single ELM crashes by introducing arbitrary seed perturbations to unstable initial conditions (with the notable exceptions of small high-frequency repetitive ELM simulations~\cite{orain2015resistive,xu2016multi})~\cite{huijsmans2015modelling}. Small differences in the chosen initial perturbations can have severe implications on the resulting dynamics and, therefore, results that depend on the amplitude and/or structure of the initial perturbations. Further, simulations that start from unstable profiles cannot answer how the plasma reached the unstable conditions in the first place.

We present for the first time non-linear MHD simulations of multiple type-I~ELM cycles. The simulated ELM repetition frequency is directly proportional to the heating source -- also an important breakthrough. Additionally, a triggering mechanism for the explosive onset of the ELM is identified and described. The simulations shown here are a first of their kind in that they repetitively reproduce realistic ELM sizes with experimentally relevant timescales. Self-consistency of the perturbations that act as initial conditions for the ELMs is achieved because the perturbations retain a characteristic structure and a non-negligible amplitude determined by the last ELM -- a feature of paramount importance for future studies regarding ELM triggering, suppression, and mitigation. Therefore, the work detailed here is an important step towards predictively studying the impact of natural type-I~ELMs and the applicability -- and robustness -- of mitigation and suppression techniques to ITER.

\section{ELM phenomenology} 
Comparisons between theory and experiment have identified ELMs~\footnote{hereafter, unless specified otherwise, ELMs refer to type-I ELMs.} as the coupling of two MHD instabilities -- the peeling mode and the ballooning mode. The peeling mode has a long wavelength ($\parallel$ to the magnetic field) and a low toroidal mode number. It is driven by the current density gradient and stabilized by the pressure gradient. Conversely, the ballooning mode is a short wavelength and high toroidal mode number instability driven by the pressure gradient, $\nabla p$, on the bad curvature side, and stabilized by large current density $j$~\cite{connor1998magnetohydrodynamic,snyder2002edge}. At the edge of H-mode plasmas with large $\nabla p$ and $j$, these instabilities couple into peeling-ballooning (PB) modes and, if the stabilising/destabilising balance between $\nabla p$ and $j$ allows, cause an ELM crash.

Experimental analyses of ELMs often include linear ideal MHD simulations probing stability with respect to PB modes at different time points. These studies almost always find the pre-ELM crash profiles to be very near a so-called peeling-ballooning stability boundary. However, it is not clear whether the ELM onset occurs exactly when the stability boundary is crossed, and what is the role of non-linear interactions on the ELM onset. Linear simulations usually ignore non-ideal effects such as resistivity as well as plasma flows, both of which are known to affect the growth rates of MHD instabilities on astrophysical and laboratory plasmas~\cite{glasser1975resistive,drake1983stabilization,diamond1985kinetic,rogers1999diamagnetic,hastie2000effect,huysmans2001modeling,velli1986resistive,swisdak2010vector,fundamenski2007relationship}. In particular, the plasma flow, primarily determined by momentum input and by the ExB velocity, is known to have an important stabilising effect on pressure-gradient-driven ballooning modes, and therefore may move the PB stability boundary~\cite{drake1983stabilization,diamond1985kinetic,rogers1999diamagnetic,hastie2000effect,huysmans2001modeling}. In the edge of H-mode plasmas, the radial electric field is set by a dominant ion diamagnetic contribution (${\sim \nabla p_i / n_i}$, where $n_i$ is the ion density) and a small ${\mathbf{v}\times\mathbf{B}}$ contribution~\cite{cavedon2017pedestal}.

The JOREK code~\cite{huysmans2007mhd,czarny2008bezier}, which solves the reduced visco-resistive single fluid MHD equations~\cite{strauss1997reduced,franck2014energy} in realistic divertor tokamak geometry, was developed in particular to study ELMs. Simulation results have already successfully captured many key characteristics of natural, triggered, and mitigated single ELM crashes in a qualitatively and quantitatively accurate manner~\cite{huysmans2009non,huijsmans2013non,orain2013non,pamela2015non,hoelzl2018insights,becoulet2014mechanism,futatani2014non,becoulet2017non}. Furthermore, it has been possible to simulate small, repetitive, high-frequency ELM crashes~\cite{orain2015resistive,pamela2015non,becoulet2017non}. Considering the stabilising effect of plasma flows (with the ion diamagnetic contribution to $E_r$~\cite{orain2013non,morales2016edge}) was key to obtain cyclical dynamics and accurate divertor heat deposition~\cite{orain2015resistive}. Simulating type-I~ELM cycles carries significant computational costs because of the need to resolve the short timescales of the ELM crash and the long timescales of the inter-ELM evolution~\cite{huijsmans2015modelling}.

\section{Type-I ELM cycles}
The starting point of the simulation is a stable and stationary post-ELM crash equilibrium reconstruction of an ASDEX-Upgrade (AUG) discharge obtained with CLISTE~\cite{mc1999analytical}. The plasma has low triangularity, high separatrix density (${n_\mathrm{sep}\sim0.4n_\mathrm{GW}}$), and no momentum input is considered. We impose heat and particle radial diffusion coefficients with an edge transport barrier together with heat and particle sources to build up a steep pressure profile. The radial diffusion coefficients and sources are static throughout the simulation time. These are used to account for physical effects beyond the scope of MHD. Namely, and inm a very simplified manner, neoclassical and anomalous transport are represented through diffusion coefficients, and heating and fuelling through the source terms. Realistic Spitzer-H\"arm parallel heat diffusion is considered and the resistivity at the plasma edge is chosen within the experimental expectation of the neoclassical resistivity. With the increasing $\nabla p$, the diamagnetic contribution to $E_r$ and the bootstrap current develop self-consistently (we consider ${\nabla p_i = \nabla p / 2}$ because the single fluid model used here does not distinguish $T_e$ and $T_i$, ). The latter is built up by considering a source term through the Sauter formula~\cite{sauter1999neoclassical,sauter2002erratum}. 

\begin{figure}[!h]
\centering
  \includegraphics[width=0.45\textwidth]{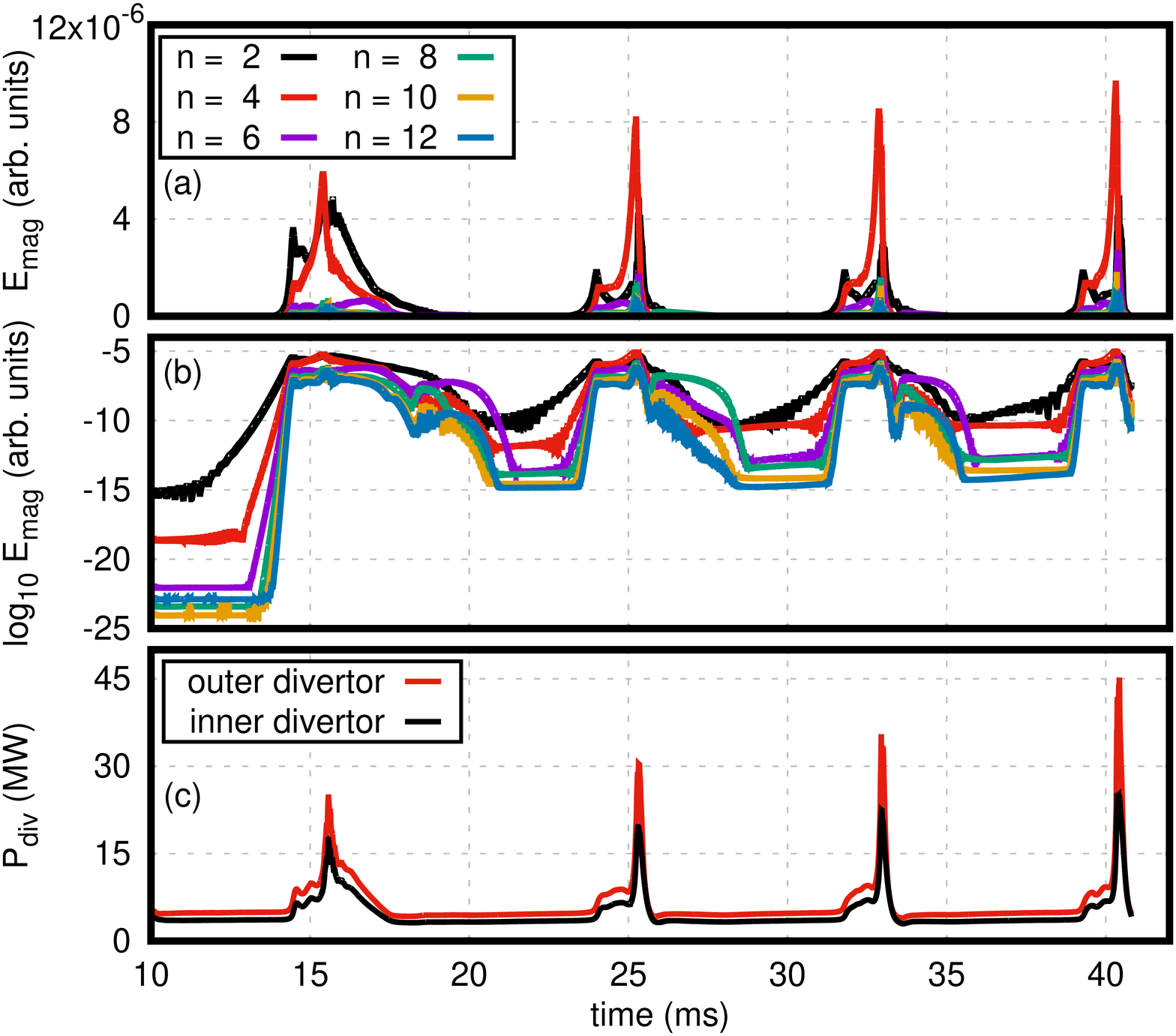}
\caption{Magnetic energies of the non-axisymmetric perturbations rising and falling at each ELM crash in linear (a) and logarithmic (b) scales. The arbitrary seed perturbations at $10~\textrm{ms}$ lead to a critically different ELM crash with respect to the next three ELMs borne out of self-consistent perturbations. (c) Power incident on the inner and outer divertor tiles in time. The outer divertor receives $\sim59\%$ of the total power during the inter-ELM phase, and $\sim51\%$ during the ELM crash.}
\label{fig:nolog-log}
\end{figure}

The plasma core, which is also part of the simulation domain, is unstable to a $2/1$ tearing mode. In order to simultaneously avoid interference between this mode with the cyclical dynamics of the ELMs and to reduce the computational cost, we include all even toroidal mode numbers between $n=0$ and $12$, i.e. simulate a half-tokamak. Nevertheless, the triggering mechanism detailed in the next section remains unchanged for a simulation with the entire toroidal mode spectrum, and the thermal energy lost in the full- and half-tokamak simulations show only a relative difference of $~\sim 2\%$. Additionally, the radial and poloidal resolution used for the present simulations is found to be properly converged. Including higher toroidal mode numbers leads to faster dynamics, but does not change the triggering mechanism or the range of dominant toroidal mode numbers. However, increasing the toroidal resolution for the full 40 ms simulation time of fig.~\ref{fig:nolog-log} is computationally not affordable for us at present. Non-axisymmetric perturbations of all the non-zero toroidal mode numbers allowed in the simulation are introduced at noise-level. Figure~\ref{fig:nolog-log}(a) and (b) show the time evolution of their magnetic energies.

As a PB stability boundary is crossed due to the simultaneously large $\nabla p$ and $j$, a low frequency ELM precursor phase begins with an $n=2$ perturbation becoming unstable, as can be seen in fig.~\ref{fig:nolog-log}(b) at $t\sim12~\mathrm{ms}$. This perturbation non-linearly drives additional modes with larger $n$ through three-wave interactions~\cite{krebs2013nonlinear}. Accordingly, during the early non-linear phase (e.g. $12$ to $13~\mathrm{ms}$ for the first ELM) the growth rate of the driven modes corresponds to the sum of the driving modes and, therefore, the highest toroidal mode number usually is the fastest growing mode. Given that the strongest non-linear coupling comes from low-n to high-n modes (because of the much higher energies of the low-n modes with respect to the energies of the high-n modes) it is not necessary to include arbitrarily more high-n toroidal modes in the simulation. The growth rate of the precursors increases with time, as expected when slowly driving the plasma across an instability boundary~\cite{callen1999growth}. The existence of such low frequency, low-n precursor activity has been observed across different tokamaks~\cite{kass1998characteristics,mink2016toroidal,mink2017nonlinear,oyama2001collapse,perez2004type,maingi2005h,kirk2014recent}. These precursors cause moderate increases in the divertor incident power, fig.~\ref{fig:nolog-log}(c), and are qualitatively similar to experimentally observed slow increases lasting $\gtrsim 1~\mathrm{ms}$ prior to the ELM~\cite{eich2003power}.

\begin{figure}[!h]
\centering
  \includegraphics[width=0.45\textwidth]{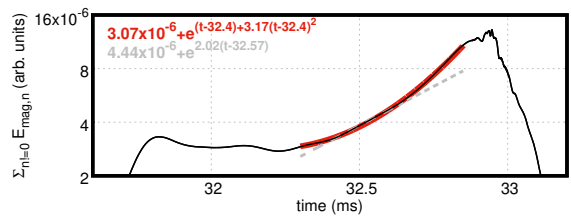}
\caption{Precursor phase and ELM crash for the third simulated ELM. The explosive onset of the ELM occurs when a phase with faster than exponential growth takes place. The sum of the magnetic energies is shown in black. The exponential and faster than exponential fitting functions are plotted in dashed grey and full red lines, respectively.}
\label{fig:prec-log}
\end{figure}

Thereafter, the $n=2$ perturbation coupled mostly with $n=4$ act together to modify the background axisymmetric plasma in sub-millisecond timescales and cause a gradual decrease of $\nabla p$ and $j$, and an even faster slowing down of the plasma flow. These timescales are shortened in simulations with higher toroidal mode numbers, but the faster slowing down of the plasma flow with respect to that of $\nabla p$ and $j$ always remains present. After this initial decrease, an explosive growth phase begins. This marks the end of the precursor phase, and the onset of the first ELM crash phase which lasts $\sim1.5~\mathrm{ms}$. The same mechanism is responsible for all of the simulated ELMs. The sum of the magnetic energies of all $n\neq0$ during the precursor and ELM crash phases is plotted against exponential and faster than exponential fitting functions in fig.~\ref{fig:prec-log}, thereby showing the explosive nature of the ELM onset. The modification of the background axisymmetric plasma due to the precursors leads to a small reduction of the energy of the perturbations (cf. fig.~\ref{fig:prec-log} from $31.8$ to $32.2~\mathrm{ms}$).

Directly after the end of the ELM crash, $\nabla p$ begins to gradually recover (which drives $j$ and $E_r$) and excites inter-ELM modes with $n$ mainly between $6$ and $8$ as seen in fig.~\ref{fig:nolog-log}(b) from roughly $18$ to $21~\mathrm{ms}$, $26$ to $27~\mathrm{ms}$, and $34$ to $35~\mathrm{ms}$. Similar inter-ELM modes, with toroidal mode numbers between $5$ and $8$, have been observed in AUG~\cite{mink2016toroidal} and KSTAR~\cite{lee2015toroidal} (the latter were simulated with JOREK~\cite{becoulet2017non}). Afterwards, the amplitudes of the non-axisymmetric perturbations become several orders of magnitude weaker than those during the ELM crash, but over up to 10 orders of magnitude stronger than their arbitrary initial amplitudes before the first ELM. The weak perturbations become destabilized again when $\nabla p$ and $j$ are large enough to simultaneously excite PB modes and overcome the stabilising effect of the plasma flow. At this point the cycle repeats itself, and there is another precursor phase followed by an ELM crash. This second ELM crash expels roughly $6\%$ of the plasma stored energy and lasts $\sim550~\mathrm{\mu s}$, which is more than twice as fast as the first ELM, which expels $\sim11\%$ of the stored energy. Due to the comparatively faster nature of the second ELM crash, with respect to the first ELM, the peak non-axisymmetric magnetic energies and the peak divertor incident power are larger for the second ELM than for the first ELM (as shown in fig.~\ref{fig:nolog-log}).

The most important difference between the first ELM and the subsequent ELMs are the seed perturbations that precede each ELM crash. For the first ELM, the seed perturbations are arbitrary as they do not hold information of the prior existence of an ELM. For the next ELMs, the seed perturbations are self-consistent with the prior existence of an ELM crash (they have a non-negligible amplitude and maintain a PB structure at all times). Therefore, the first time the PB stability boundary is crossed the seed perturbations require more time to affect the background plasma with respect to the subsequent times that it is crossed (as seen in fig.~\ref{fig:nolog-log}(b)). Consequently, the pressure directly before the first ELM builds up to larger values than before the subsequent ELMs, and the expelled thermal energy is larger for the first ELM and results in a longer ELM crash. This behavior is reminiscent of ``giant'' ELMs that expel larger amounts of thermal energy and have a longer duration than regular type-I~ELMs. These appear after extended ELM-free phases, during which the seed perturbations may become weaker and lose their PB mode structure ~\cite{jackson1991regime,nave1997overview,chankin2003influence,huysmans1998identification}. Because of the discrepancies between the first (giant) and all the subsequent ELMs, in the following section we will focus on the latter to describe the triggering mechanism for the explosive onset of the ELM crash. In reality, the remnant MHD activity after an ELM crash may interact with (or become affected by) micro-turbulence during the inter-ELM period (depending on their respective spatial scales). The seed perturbations then result from both types of activity. However, such dynamical effects cannot be addressed with the present set-up and the seed perturbations are comprised exclusively of the remnant MHD activity from the last ELM crash.

It is worth pointing out that comparing the first and the subsequent ELMs is somewhat flawed as the density and temperature pre-ELM profiles are slightly different (not only due to the time required for the seed perturbations to grow to observable amplitudes). To produce a more robust comparison between ELMs with arbitrary and with self-consistent seed perturbations, we have performed an additional test. We eliminate the non-axisymmetric perturbations from the simulation after the second ELM crash (at $28.4~\mathrm{ms}$) and immediately introduce perturbations again at noise level -- like was done before the first ELM crash. This additional simulation (not shown) requires more time for the seed perturbations to grow and affect the background plasma and, therefore, the pre-ELM pressure is higher than for its counterpart with self-consistent seed perturbations (the third ELM crash from fig.~\ref{fig:nolog-log}). This further evidences the importance of self-consistent seed perturbations for ELM simulations.

\begin{figure}[!h]
\centering
  \includegraphics[width=0.45\textwidth]{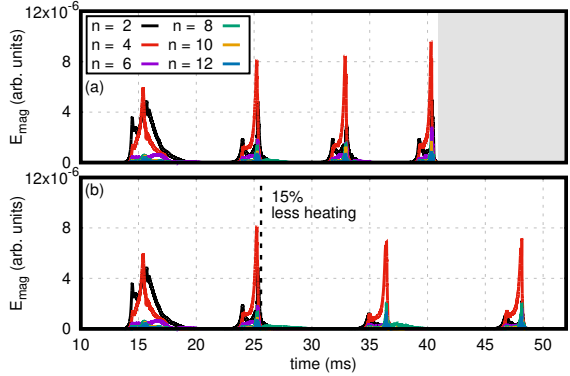}
\caption{Magnetic energies of the non-axisymmetric perturbations for nominal (a) and $85\%$ nominal (b) heating. The ELM repetition frequency for (a) is $f_{ELM}\approx120~Hz$, and it is reduced to $f_{ELM}\approx87~Hz$. The nominal heating simulation is only performed until $40.9~ms$.}
\label{fig:heating-nolog}
\end{figure}

The imposed diffusion coefficients, the applied heating power, and the particle source govern the timescale at which $\nabla p$ grows. The pedestal build-up in reality results from dynamic anomalous and neoclassical transport, applied heating power and fuelling including neutrals recycling. Realistically accounting for such dynamical effects, in order to produce predictive modelling, goes beyond the scope of this investigation. Nevertheless, in order to ensure that the simulated macroscopic instabilities are type-I ELMs, we investigate how they respond to changes in the injected heating power. This scan can be seen as modifying the build-up time scale of the pedestal. In doing so, we observe a direct dependency of ELM frequency with heating power, therefore bolstering the argument that type-I~ELMs are simulated. Reducing the heating power by $15\%$ leads to a lower ELM repetition frequency, as shown in fig.~\ref{fig:heating-nolog}. A thorough heating and fuelling scan with a more realistic model for the pedestal evolution is envisioned as future work. 

\section{ELM triggering mechanism}

By analyzing the simulation results we find that the influence of the precursors on the background axisymmetric plasma is responsible for the explosive ELM onset. The underlying mechanism relies on the existence of reconnection of magnetic field lines (taking place due to the non-zero resistivity) and on a separation of timescales between the responses of $\nabla p$ and $E_r$ to the enhanced transport by stochastic magnetic topology. 

\begin{figure}[!h]
\centering
  \includegraphics[width=0.48\textwidth]{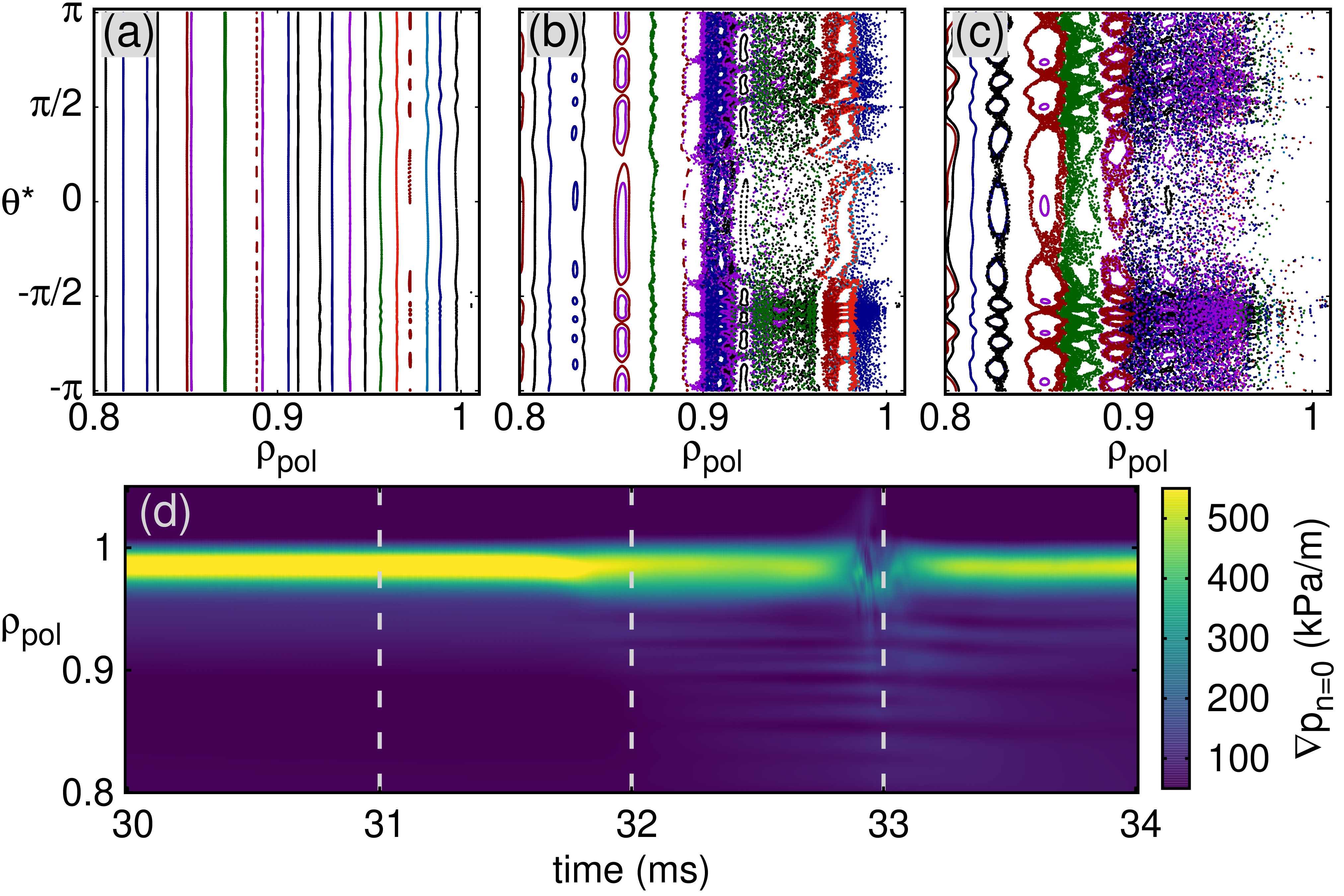}
\caption{Precursor phase and ELM crash showing (a)-(c) Poincar\'e plots of the magnetic field lines at $31$, $32$, and $33~\mathrm{ms}$ respectively, and (d) time-evolving outboard midplane toroidally averaged pressure gradient. Precursor activity lasting roughly $1~\mathrm{ms}$ starts at $\sim31.8~\mathrm{ms}$. We use the radial coordinate, $\rho_{pol}=\sqrt{\psi_N}$ where $\psi_N$ is the normalized poloidal magnetic flux equal to $0$ in the magnetic axis and $1$ at the separatrix, and the poloidal coordinate, $\theta^*$ equal to $0$ at the outboard midplane and $-\pi/2$ at the magnetic x-point.}
\label{fig:poincare}
\end{figure}

As the precursor amplitude becomes large enough ($\delta n_e/n_e\sim1$), the edge magnetic field starts to ergodize. Figure~\ref{fig:poincare}(a) shows magnetic field lines inside the separatrix closing in at the same flux surface where they started at $31~\mathrm{ms}$. One millisecond later, fig.~\ref{fig:poincare}(b) shows field lines that no longer necessarily arrive at the same flux surface where they started because axisymmetry is broken by the strong precursor activity.

\begin{figure}[!h]
\centering
  \includegraphics[width=0.48\textwidth]{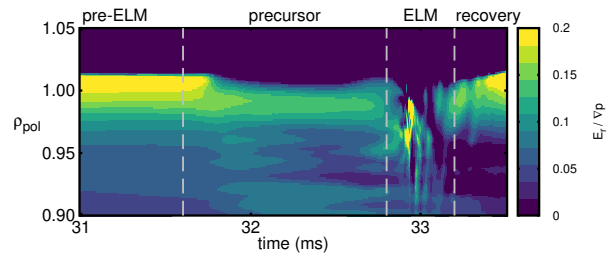}
  \caption{Time evolution of the outboard midplane axisymmetric ratio $E_r/\nabla p$ in the pedestal region. The ratio shows the balance between the stabilising $E_r$ and the destabilising $\nabla p$. It steadily decreases (notably around the maximum pressure gradient region: $\rho_\mathrm{pol}\sim0.99$) when the precursor phase begins at $\sim31.6~\mathrm{ms}$, therefore indicating increasingly unstable conditions which set the stage for the explosive ELM onset. The ratio increases again when the ELM crash ends at $\sim33.2~\mathrm{ms}$.}
\label{fig:diff}
\end{figure}

The non-axisymmetric magnetic topology during the precursor phase connects flux surfaces at different radial positions and, therefore, drastically increases diffusive parallel heat transport. This widens and rapidly flattens the temperature gradient across ${\rho_{pol}\approx[0.96-1.00]}$. Therefore causing $\nabla p$ to change in the same manner, clearly shown in fig.~\ref{fig:poincare}(d). Since stochastic transport affects temperature gradients faster than it affects density, $\nabla p$ decreases faster than density does~\cite{Wesson:1427009}. Additionally, $E_r$ decreases in a faster time scale than $\nabla p$, as clearly evidenced in fig.~\ref{fig:diff}. The second destabilising term, $j$, changes even slower than $\nabla p$ through current diffusion. We reiterate that the precursor timescales are faster when higher toroidal modes are considered and therefore we do not venture to compare the temporal dynamics to low frequency low-n precursors observed in experiment.

At first glance, the changes to the plasma caused by the precursors may seem stabilising. Namely, the decrease of $\nabla p$ and $j$ in the pedestal are, from the linear ideal MHD picture, stabilising effects. However, the stabilising effect of $E_r$ decreases faster than the destabilising effect of $\nabla p$ as shown in fig.~\ref{fig:diff} where four distinct phases can be observed. The progressively smaller ratio $E_r/\nabla p$ means that the existing PB modes in the pedestal become less restricted by the stabilising effect of $E_r$ and may grow progressively faster (i.e. explosively) until they cause the ELM crash. At the same time, a localised increase of $\nabla p$ also resulting from the precursor activity can locally drive the plasma further into the unstable regime. These effects become self-amplifying~\cite{wilson2004explosive} and the total magnetic energy of the perturbations rises explosively. During the non-linear ELM onset, both effects play an important role.

The initial pre-ELM phase sustains a roughly constant $E_r/\nabla p$. The faster slowing down of the plasma flows with respect to $\nabla p$ marks the beginning of the precursor phase. During this phase $E_r/\nabla p$ quickly decrease (fig.~\ref{fig:diff}), thereby leading to the explosive ELM onset (fig.~\ref{fig:prec-log}) until it abruptly ends with the ELM crash at $\sim32.8~\mathrm{ms}$. The changes to the axisymmetric background during the precursor phase triggers PB modes to grow explosively and couple between one-another while at the same time making the ergodic region penetrate further inwards, as evidenced by the change from fig.~\ref{fig:poincare}(b) to (c). The ELM crash phase features losses due to the increasingly ergodic magnetic topology and from convective transport occurring in sub-millisecond timescales directly comparable to experimental observations~\cite{iter1999mhd}. Finally, the recovery phase takes place once the ELM crash is concluded. During this phase $E_r/\nabla p$ returns to the pre-ELM state and the magnetic topology becomes close to axisymmetric again. Even though $E_r/\nabla p$ recovers in a sub-millisecond timescale after the crash, $E_r$ and $\nabla p$ individually require roughly $7~\mathrm{ms}$ to return to the pre-ELM state.

\section{Discussion and conclusions}
We present, for the first time, simulations of realistic type-I~ELM cycles in diverted tokamak geometry. Important differences in the modelled ELM crash dynamics (notably their size and duration) are observed with different initial seed perturbations. The first simulated ELM, with arbitrary seed perturbations, results in a longer ELM crash with more energy lost when compared to the subsequent ELM crashes with self-consistent seed perturbations. Since the seed perturbation depend on the non-linear dynamics of the previous ELM, we conclude that in order to use the present numerical tools to predictively assess the consequences of natural ELMs, or the applicability of existing ELM mitigation and suppression techniques to future tokamaks, it is necessary to model full ELM cycles. 

From the simulation results we identify a non-linear electromagnetic triggering mechanism for the explosive ELM onset. During the precursor phase, an increasingly stochastic magnetic topology causes a decrease of $\nabla p$ and $j$ with an even faster slowing down of the plasma flows. Consequently, the stabilising effect of the plasma flows is rapidly lost and prompts an explosive ELM onset. 

Given that a single fluid temperature was considered, the parallel heat transport resulting from the stochastic magnetic topology does not account for the separation in electron and ion timescales. We expect only the precursor phase duration to be modified as a result. Additionally, in experiments the inter-ELM evolution shows separate timescales between $T_e$ and $T_i$, which affects the diamagnetic contribution to $E_r$~\cite{cavedon2017pedestal}. Therefore, separating the electron and ion temperature evolution is envisioned for future work. The diffusive transport of particles, and the ion and electron heat flux in the experiment is not determined by static diffusion coefficients like we have assumed here for simplicity. Future investigations into more accurate pedestal evolution are also of interest as they may shed light onto other inter-ELM modes and high mode number precursors.

Simulations with higher toroidal harmonics (all even modes until $n=20$), or with the entire toroidal mode spectrum ($n=0,1,2,...,12$), feature the same ranges of dominant toroidal mode numbers and same triggering mechanism with explosive onset, albeit with shorter precursor phases. Nonetheless, the observed non-linear triggering mechanism is robust to changes in the chosen toroidal mode numbers and to variations of the imposed inter-ELM evolution, i.e. changes in heating power. In general, the simulated ELM crashes and precursors show characteristics that are qualitatively consistent with observed ranges of toroidal mode numbers, ELM sizes and duration, and divertor heat loads, to name a few. Finally, the ELM repetition frequency of the simulated ELMs shows a direct dependency to the applied heating power, as expected for type-I~ELMs.

\section*{Acknowledgements}

This work has been carried out within the framework of the EUROfusion Consortium and has received funding from the Euratom research and training program 2014-2018 and 2019-2020 under grant agreement No 633053. The views and opinions expressed herein do not necessarily reflect those of the European Commission. This work used the MARCONI computer at CINECA under projects AUGJOR and ELM-UK.

\bibliography{references}

\end{document}